\documentclass[prl,onecolumn,superscriptaddress,amsmath,amssymb]{revtex4}
\usepackage{color,graphicx,amsmath,upgreek}

\begin{document}

\title{Efficient, Compact and Low Loss Thermo-Optic Phase Shifter in Silicon}

\author{Nicholas C. Harris}
\affiliation{Department of Electrical Engineering and Computer Science, Massachusetts Institute of Technology, Cambridge, MA 02139, USA}
\author{Yangjin Ma}
\affiliation{Department of Electrical and Computer Engineering, University of Delaware, Newark DE 19716, USA}
\author{Jacob Mower}
\affiliation{Department of Electrical Engineering and Computer Science, Massachusetts Institute of Technology, Cambridge, MA 02139, USA}
\author{Tom Baehr-Jones}
\affiliation{Department of Electrical and Computer Engineering, University of Delaware, Newark DE 19716, USA}
\author{Dirk Englund}
\affiliation{Department of Electrical Engineering and Computer Science, Massachusetts Institute of Technology, Cambridge, MA 02139, USA}
\author{Michael Hochberg}
\affiliation{Department of Electrical and Computer Engineering, University of Delaware, Newark DE 19716, USA}
\author{Christophe Galland}\email{chris.galland@free.fr}
\affiliation{\'{E}cole Polytechnique F\'{e}d\'{e}rale de Lausanne (EPFL), CH-1015 Lausanne, Switzerland}

\begin{abstract}
We design a resistive heater optimized for efficient and low-loss optical phase modulation in a silicon-on-insulator (SOI) waveguide and characterize the fabricated devices. Modulation is achieved by flowing current perpendicular to a new ridge waveguide geometry. The resistance profile is engineered using different dopant concentrations to obtain localized heat generation and maximize the overlap between the optical mode and the high temperature regions of the structure, while simultaneously minimizing optical loss due to free-carrier absorption. A 61.6 $\mu$m long phase shifter was fabricated in a CMOS process with oxide cladding and two metal layers. The device features a phase-shifting efficiency of 24.77 $\pm$ 0.43 mW$/\pi$ and a -3 dB modulation bandwidth of 130.0 $\pm$ 5.59 kHz; the insertion loss measured for 21 devices across an 8-inch wafer was only 0.23 $\pm$ 0.13 dB. Considering the prospect of densely integrated photonic circuits, we also quantify the separation necessary to isolate thermo-optic devices in the standard 220 nm SOI platform.
\end{abstract}
\maketitle

\section{Introduction}
The silicon-on-insulator (SOI) material platform has received much recent attention for its capability to support scalable and inexpensive photonic integrated systems-on-chip. While the field has mainly targeted the telecommunication and data-interconnect industry, new applications such as phased antenna arrays \cite{Sun:2014kl,Kwong:2014kd} and quantum photonic circuits \cite{Bonneau:2012} are attracting increased interest. Two mechanisms are most commonly used for effecting a change of silicon's index of refraction: free-carrier plasma dispersion and the thermo-optic effect \cite{Jalali:2006td}. The plasma dispersion effect has been widely leveraged to realize modulation at rates above 10 GHz \cite{BaehrJones:2012ta,Novack:2013fj}. However, the intrinsic optical loss due to free-carrier absorption makes this approach unsuitable for a number of emerging applications including integrated quantum optics. First, these modulators exhibit large passive insertion losses that make their use in large scale circuits rapidly prohibitive. Even more deleterious are the intrinsic dynamic losses which prevent pure phase modulation, as is required for tuning inteferometers without degradation of interference visilibity or tuning resonators without degradation of their quality factor. Other schemes such as coherent homodyne and heterodyne detection also benefit from lossless phase modulation.

As a result of the relatively large thermo-optic coefficient of silicon near 300 Kelvin and at wavelengths near 1550 nm, $dn/dT=1.86 \times 10^{-4}$ K$^{-1}$ \cite{Rendina:1992} where $n$ is the refractive index and $T$ is temperature in Kelvin, thermal effects have been successfully used to tune and stabilize ring resonators \cite{Padmaraju:12,Fegadolli:2014th} and interferometric switches \cite{Watts:2013vx, VanCampenhout:2010uw, Fang:2011gp, Chu:2005vg, Song:2008wi}. Yet, when considering, for example, the development of large-scale quantum photonic circuits based on reconfigurable quantum gates \cite{Shadbolt:2011vo}, previously demonstrated thermo-optic phase shifters are quite long (preventing dense intregration), have notable insertion loss, or are not implemented with a standard silicon dioxide cladding used in complementary metal-oxide-semiconductor (CMOS) processes for passivation and metal layer fabrication, as shown in Table 1. While phase shifters operating at high rates and with low power requirements are desirable, these characteristics are difficult to achieve simultaneously. This is made clear by expressing power consumption in terms of speed, as P$_\pi = \frac{H}{\tau}$ $\Delta$T$_\pi$ where $H$ is heat capacity, $\Delta$ T$_\pi$ and P$_\pi$ are the change in temperature and power dissipation required to achieve $\pi$ phase shift, and $\tau$ is the thermal time constant \cite{Watts:2013vx}. This relation expresses the fact that a fast and low-power device, characterized by a small P$\pi \cdot \tau$ product, has to be as small as possible (small $H$)---but this restricts the length of the modulation region and therefore asks for large $\Delta T_\pi$, leading to the unavoidable trade-off between speed and power consumption. In many of the designs in Table 1, inneficiency and excess length \cite{Xia:2004vq} may be attributed to weak localization of heat to the waveguiding region while extreme efficiency levels \cite{Fang:2011gp} are largely due to oxide undercuts and removal of cladding. Here, we demonstrate an ultra-low loss thermo-optic phase shifter in a process with oxide cladding that is 61.6 $\mu$m long with a P$_\pi$ of 24.77 $\pm$ 0.43 mW, where P$_\pi$ is defined as the power required to achieve $\pi$ radians phase shift, and a -3 dB bandwidth of 130.0 $\pm$ 5.59 kHz. Our device operates with a V$_\pi$ of 4.36 V; more than a factor of two lower than the most competitive compact waveguide-integrated thermo-optic phase shifters \cite{Watts:2013vx} that benefit from the removal of the top oxide cladding. We also quantify the separation necessary to isolate thermo-optic devices in standard 220 nm SOI.

\begin{table}[h!]
  \centering
  \caption{Summary of recent thermo-optic waveguide phase shifter parameters where $L$ is the total heater length, V$_\pi$ and P$_\pi$ are the applied voltage and power necessary to reach $\pi$ radians of phase shift, respectively, and $\tau$ is the limiting rise or fall time constant. In results where $\tau$ is not reported, the single-pole approximation $\tau=\frac{0.35}{f_3dB}$ is used to convert between metrics.}
  \begin{tabular}{l  l  l  l  l  l  l  l  l}
    \hline
                            & Material     & Cladding  & $L$ ($\upmu$ m) & Loss (dB) & V$_\pi$ (V) & P$_\pi$ (mW)  & $\tau$ ($\mu$s) & P$_\pi \cdot \tau$ (mW$\cdot\mu$s) \\ \hline
Here                        & SOI          & SiO$_{2}$ & 61.6            & 0.23 dB   & 4.36        & 24.77         & 2.69            & 66.9               \\
\cite{Watts:2013vx}         & Si           & Air       & $>$9.42         & 0.5 dB    & 11.93       & 12.7          & 2.4             & 30.5               \\
\cite{Fang:2011gp}          & TiN in SOI   & Air       & 1000            & 0.3 dB    & 0.86        & 0.49          & 144             & 70.5               \\
\cite{VanCampenhout:2010uw} & NiSi in SOI  & Air       & 200             & 5 dB      & 1           & 20            & 2.8             & 56                 \\
\cite{Espinola:2003go}      & Cr-Au in SOI & SiO$_{2}$ & 700             & 32 dB     & 1.66        & 46            & 3.5             & 160                \\
\cite{Song:2008wi}          & Ti in SOI    & Air       & 100             & 8 dB      & 13.3        & 10.6          & 34.9            & 370                \\
\cite{Xia:2004vq}           & Metal in SOI & SiO$_{2}$ & 2500            & $<$ 12 dB & -           & 235           & 60              & 14100              \\

  \end{tabular}
\end{table}

\section{Device geometry and fabrication}
Our optimized thermo-optic phase shifter (Fig. 1 (a) and (b)) was fabricated in the OpSIS process on a SOI wafer with a 220 nm thick top silicon layer \cite{baehr201225,Galland:2013je}. Two levels of boron doping by ion implantation were used, with peak concentrations of 1.7$\cdot$10$^{20}$  cm$^{-3}$ for $p$++ and 7$\cdot$10$^{17}$ cm$^{-3}$ for $p$. Two aluminum routing layers were used; the top layer functioned as an electrical probe pad layer and as a signal routing layer while the bottom layer was used for signal routing. The metal contacting region of the phase shifter was connected to the ridge waveguide using 800 nm wide channels defined in a partially etched 90 nm thick silicon slab with both $p$ and $p$++ implantation, as shown in Fig. 1(b). The measured sheet resistances were 136 $\Omega$ for the $p$++ -doped 90 nm thick Si layer, 13.6 k$\Omega$ for the $p$-doped 90 nm thick layer and 3.87 k$\Omega$ for the $p$-doped 220 nm thick layer.

\begin{figure}[h!]
  \centering 
  \includegraphics[width=18cm]{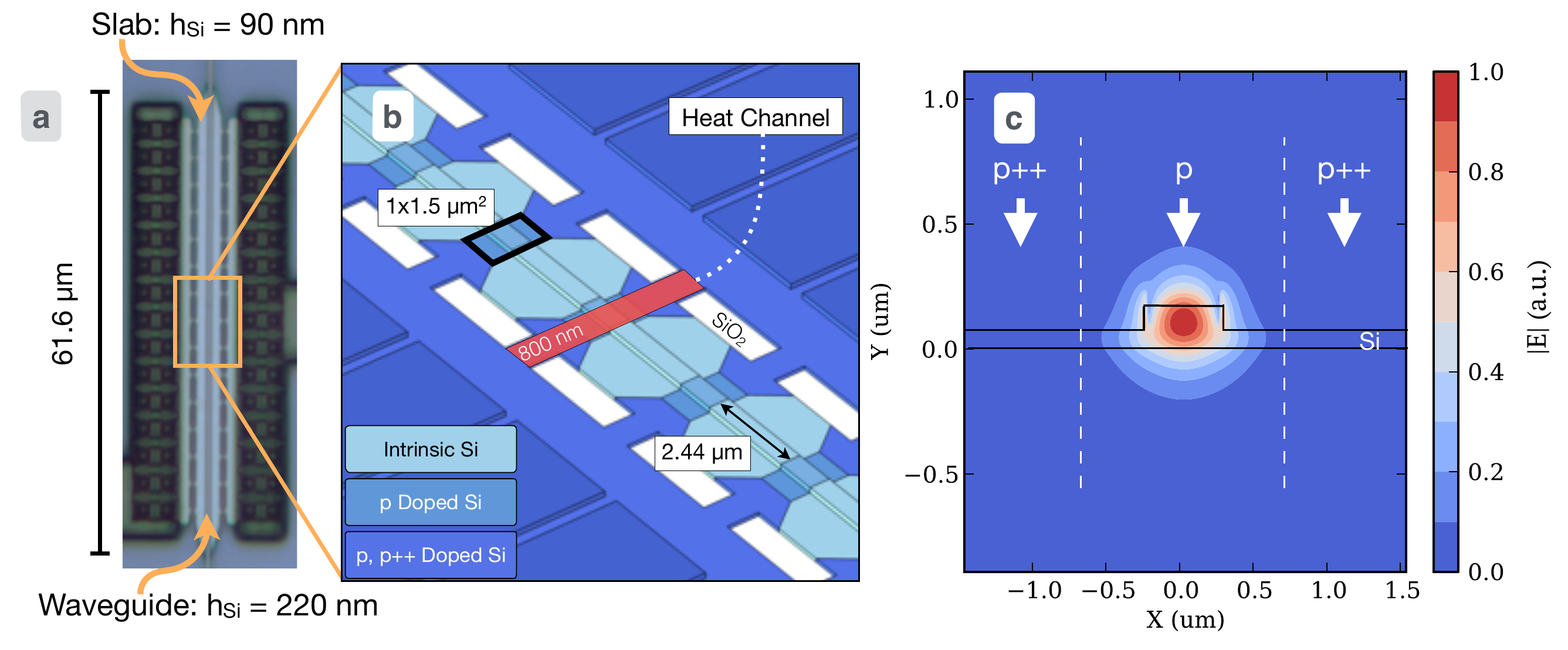}
  \caption
  {
  (a) Optical micrograph of the structure with the vias connecting the lowest metal layer and the doped silicon clearly visible. $h_{Si}$ denotes silicon layer thickness. (b) Perspective view of the phase shifter with annotations for relevant dimensions. (c) Doping profile along the cross section marked red in (b), overlapped with the simulated amplitude of the horizontal component of the electric field.
  }
\end{figure}

The design of the device largely proceeds from three principles. Overlap between the silicon-guided optical mode and the thermal profile should be maximized and heat propagation and optical loss should be minimized. By $p$-doping only the 1.0 $\upmu$m wide transverse waveguide section and $p$ and $p$++ doping elsewhere, heat can be generated in a small region with large optical mode overlap. Since the thermal conductivity of SiO$_2$ is two orders of magnitude smaller than that of silicon, the 800 nm wide channels in Fig. 1(b) connecting the contact region to the ridge waveguide efficiently restrict the outward propagation of heat. Sufficient clearance between the guiding region and the $p$++ -doped region as well as overlapping dopants with the optical mode only every 2.44 $\upmu$m avoids excess losses to free-carrier absorption, as shown in Fig. 1 (c). Tapered spot size converters allow for an adiabatic transition between the single-mode channel waveguide and ridge waveguide, preventing the excitation of the higher-order modes supported in the latter. This ensures low-loss transition between channel and ridge waveguides at the input and output of the modulator. By adding or removing unit cells corresponding to tiled thermal channel sections, it is possible to achieve a desired device resistance and operating voltage while independently choosing its length.

\section{Device Simulations}
To confirm this localization near the waveguide, we simulated the voltage and temperature fields using the COMSOL Multiphysics finite-element solver. We set room temperature boundary conditions below the buried oxide layer and 10 $\upmu$m above the top oxide cladding and used previously reported thermal \cite{Liu:2004ck,Touloukian:1970tb,Kleiner:1996vg} and electrical conductivities for the silicon layers. Since we were interested in the intrinsic device performance, we did not include the metal contacts in the simulation space.

\begin{figure}[h!]
  \centering
  \includegraphics[width=15cm]{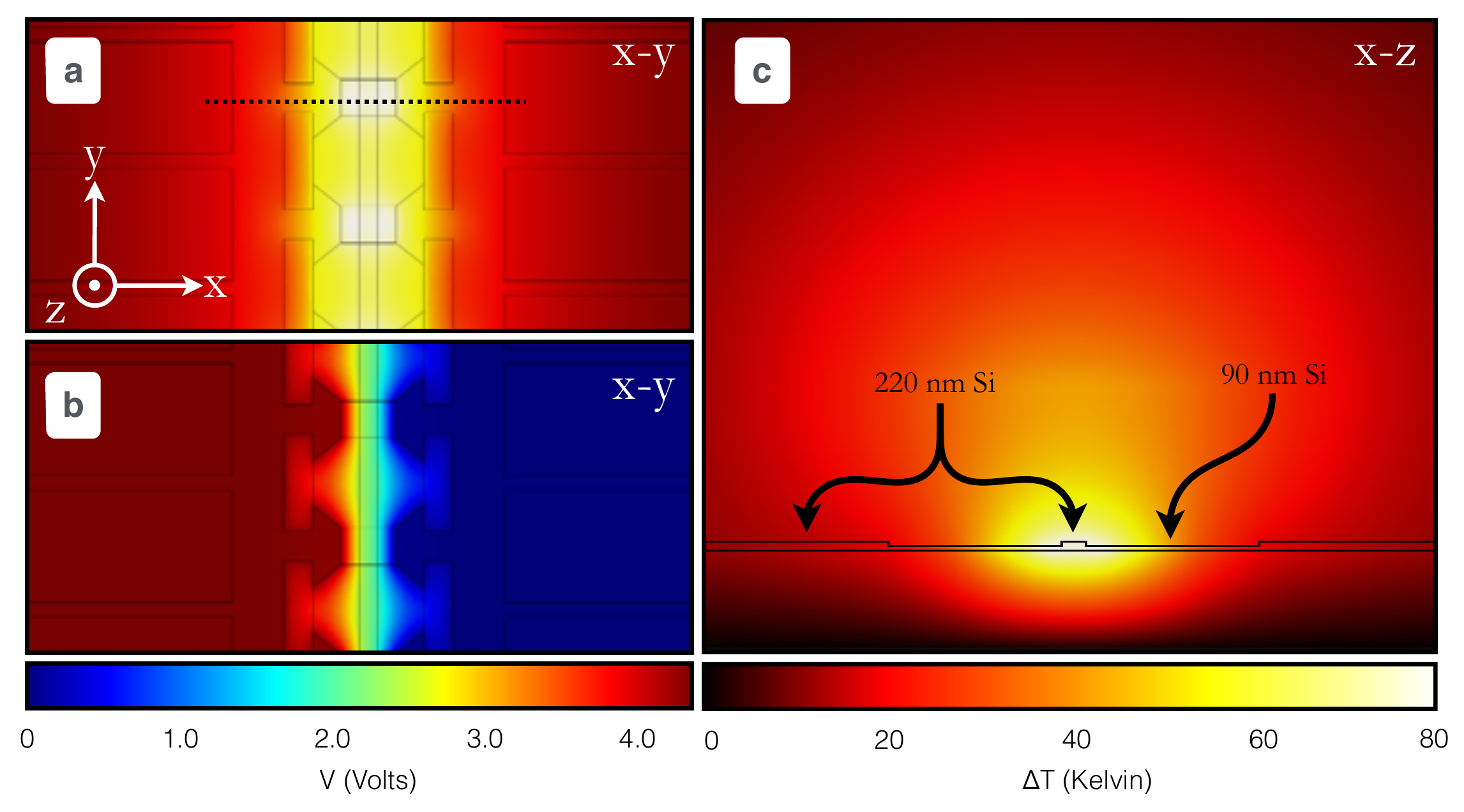}
  \caption
  {
    Planar projections of temperature and voltage distributions from a three-dimensional simulation. (a) and (b) Show the highly localized temperature and voltage distributions, respectively, for an applied voltage corresponding to $\pi$ phase shift. (c) Temperature distribution with device outline overlay for the dotted-line cross-section in (a). (a) and (c) share the same temperature color bar. The voltage is dropped almost exclusively across the thermal channel.
  }
\end{figure}

The device was simulated with the bias voltage set to 4.36 V, corresponding to the measured average value of V$_\pi$. Fig. 2(a) and (b) show the strong localization of the temperature and voltage drop near the waveguide as a consequence of the narrow thermal channels and dopant configuration employed here. The change in phase as a function of temperature can be expressed as $\Delta\Phi=\frac{2\pi L}{\lambda_0}\frac{dn}{dT}\Delta T$, where $L$ is the device length, $\lambda_0$ is the free-space wavelength, $\frac{dn}{dT}$ is the thermo-optic coefficient and $\Delta T$ is the change in temperature. This can be approximated as $\Delta\Phi \simeq 2.4\pi \cdot 10^{-4} \times \Delta T \cdot L$ for the case where $\lambda_0$ is 1550 nm and $\frac{dn}{dT}$ is taken as the room temperature thermo-optic coefficient of silicon. Fig. 2(c) shows the temperature distribution in the cladding which is largely surrounding the waveguiding region. We performed the same simulation for the case with air cladding, rather than oxide, revealing that the phase delay could increase by as much as 25\% for the same applied potential.

\section{Device Characterization}
\subsection{Thermo-optic phase shifter}
To characterize the phase shift as a function of power dissipation, the thermo-optic modualtor was fabricated as part of one arm of an unbalanced MZI, with a measured free spectral range of 6.4 nm (Fig. 3), composed of two low-loss multi-mode interferometer (MMI) y-junctions \cite{Zhang:2013tm}. We coupled light on chip using grating couplers \cite{Li:2013} and performed a spectral sweep between 1520 nm and 1570 nm for each applied voltage and power dissipation level. These spectra were then fit to a sinusoid to extract the phase shift with respect to the unbiased spectra. We plot the results in Fig. 3 (e), from which we obtain a $P_\pi$ of 24.77 $\pm$ 0.43 mW. This corresponds to $V_\pi = 4.36$ V given the device resistance of 769.00 $\pm$ 1.24 $\Omega$.

\begin{figure}[h!]
  \centering
  \includegraphics[width=16cm]{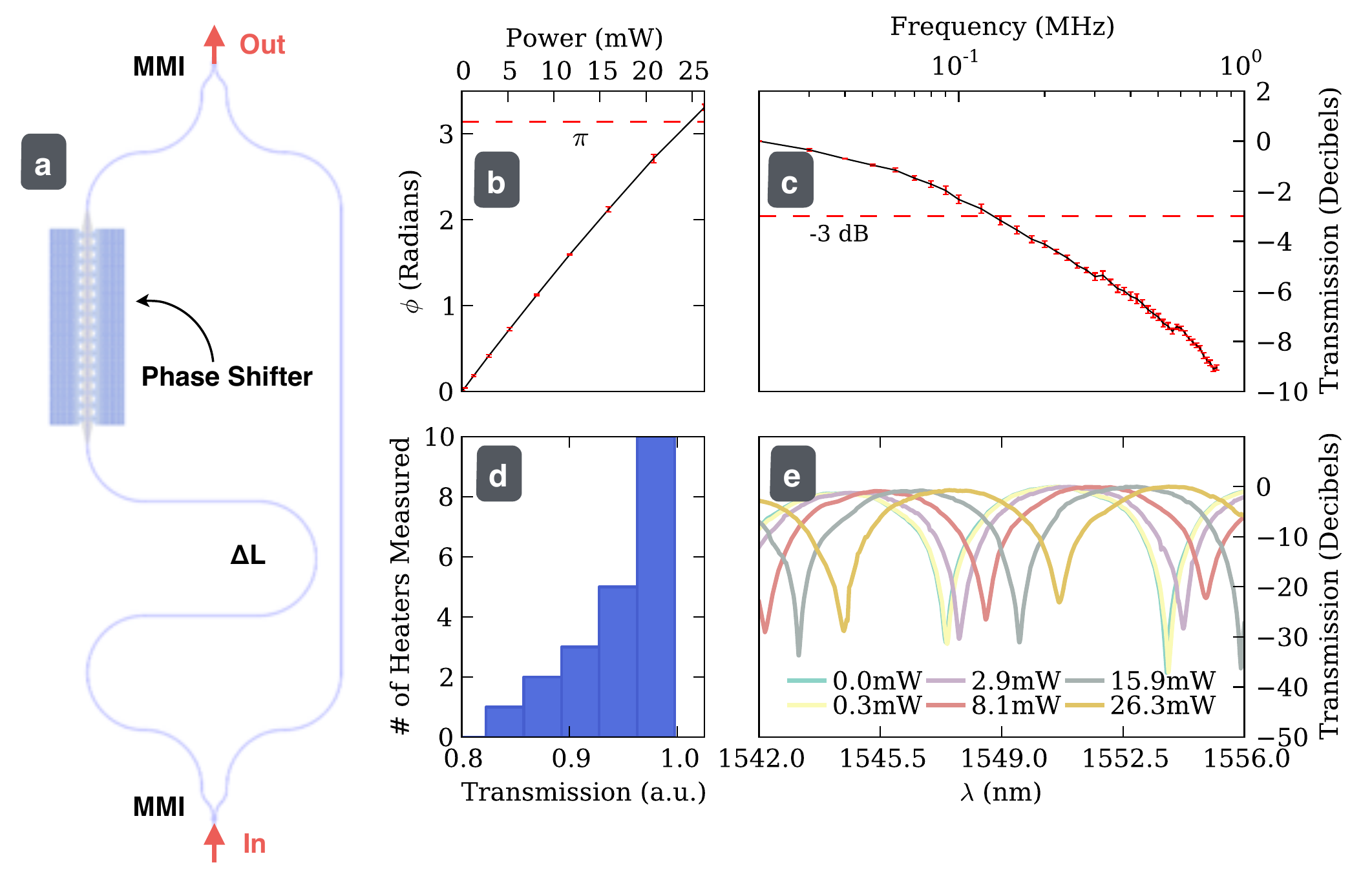}
  \caption
  {
    (a) Annotated test structure layout including MZI with path imbalance and the thermo-optic phase shifter, (b) average phase shift versus dissipated power for three devices, (c) average response of the MZI to pure sinusoids of various frequencies for four devices with a dotted line labeling the -3 dB level, (d) histogram of the transmission through the heater for 21 devices on the same wafer, (e) MZI spectra for various power dissipation levels.
  }
\end{figure}

A Stanford Research Systems lock-in amplifier was used to measure the bandwidth of the thermo-optic phase shifter. The frequency of the sinusoidal output signal was swept from 20 kHz to 800 kHz and the amplitude response was recorded at each step. The -3 dB bandwidth of the the phase shifter was measured to be 130.0 $\pm$ 5.59 kHz, as shown in Fig. 3 (e). Adjacent grating couplers were measured on each of the 21 dies and their transmission spectra was recorded. The spectra were then normalized to the grating coupler spectra, MMI insertion loss and waveguide propagation loss yielding a phase shifter insertion loss of 0.23 $\pm$ 0.13 dB. The insertion losses of 21 devices measured across an 8-inch SOI wafer are summarized in the histogram of Fig. 3 (d). The low static loss of our phase shifter enables us to achieve the deep extinction shown under both passive and active operation.

\subsection{Thermal decay test structure}
Thermal diffusion can adversely affect the performance of adjacent photonic components, which is an important constraint when designing densely integrated large-scale photonic circuits. To manage this thermal cross-talk, it is essential to quantify the necessary separation between thermo-optic devices and other phase-sensitive components.

\begin{figure}[h!]
  \centering
  \includegraphics[width=18cm]{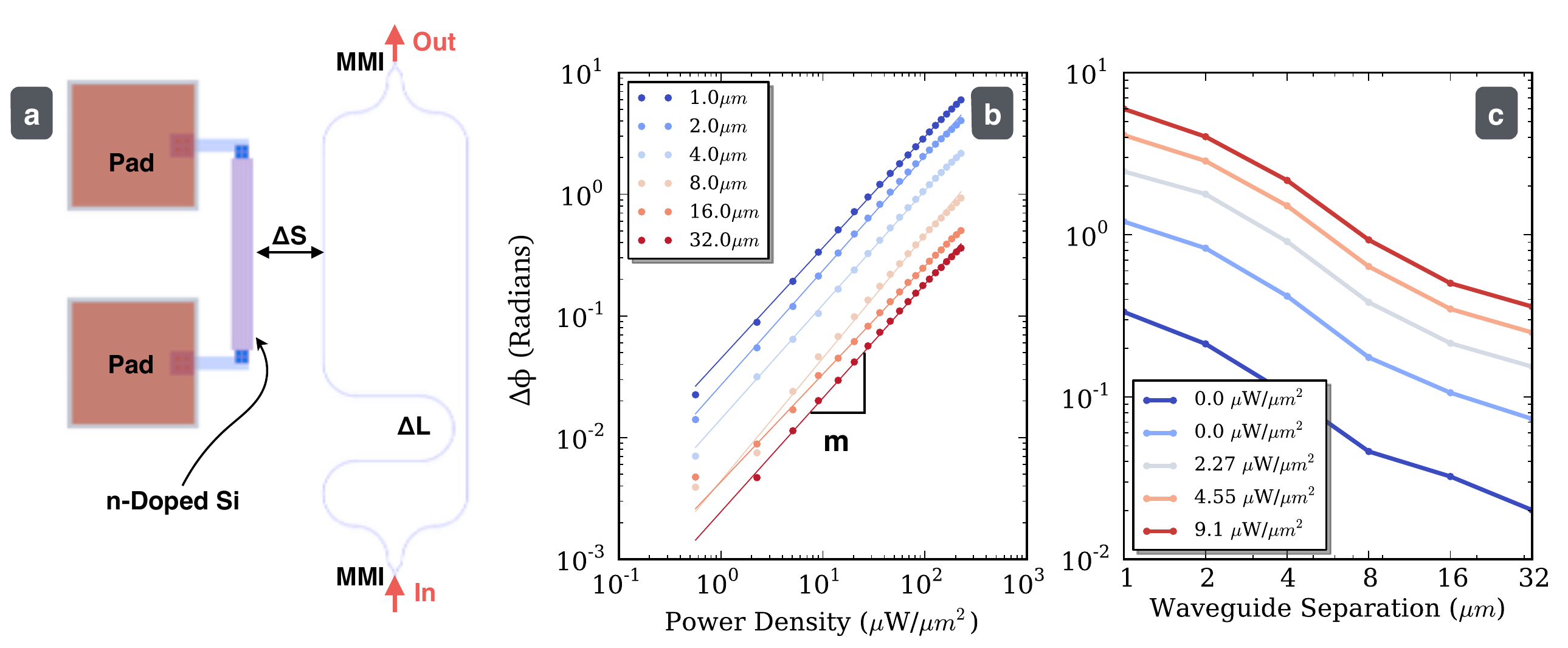}
  \caption
  {
    (a) Annotated layout for the test structure used to probe the temperature distribution within the SOI wafer. The arms of the MZI were designed to have a large separation to avoid a phase shift in both simultaneously. (b) Phase shift imparted on left-most MZI path as a function of the waveguide to heater separation for various power dissipation levels. Power law fit lines are also shown. (c) Phase shift imparted on MZI left-most arm as a function of dissipated power for various waveguide to heater separation levels.
  }
\end{figure}

We used a passive unbalanced MZI to probe the decay of heat generated by an $n$-doped (phosphorus $5\cdot 10^{-3}$ cm$^{-3}$) resistor, measuring 503.52 $\pm$ 0.09 $\Omega$, running parallel with the MZI at a distance $\Delta S$ (Fig. 4a). Six structures with $\Delta S =$ 1, 2, 4, 8, 16 and 32 $\upmu$m were tested. The phase shift was measured for each of the six structures at various bias levels, resuling in the data shown in Fig. 4 (b) and (c). The decay of the induced phase shift $\Delta\Phi$ with power density is linear, as shown in Fig. 4(b), and can be fit to a power law $\Delta\Phi = a_{0}\rho^{m}$ where $m$ is 0.937 $\pm$ 0.015. Fig. 4 (c) can be used to provide an indication of how far away waveguiding elements must be placed in order to achieve a desired isolation level.

\section{Conclusion}
We demonstrate a compact thermo-optic phase shifter that is 61.6 $\upmu$m long with a P$_\pi$ of 24.77 $\pm$ 0.43 mW and a -3 dB bandwidth of 130.0 $\pm$ 5.59 kHz. The propagation loss in the device is quite low at 0.23 $\pm$ 0.13 dB and is due to the overlap of the optical mode with the boron-doped silicon and mode conversion between the ridge and rib waveguide geometries. This new thermo-optic phase shifter design enables precise targeting of power dissipation and heat localization, resulting in low thermal crosstalk and high efficiency. We also characterized the thermal decay characteristics of heaters based on resistive, doped silicon in SOI.

\section{Acknowledgments} We would like to acknowledge AFOSR STTR grant number FA9550–12-C-0038. The authors would like to thank Gernot Pomrenke, of AFOSR, for his support of the OpSIS effort, through both a PECASE award (FA9550-13-1-0027) and ongoing funding for OpSIS (FA9550-10-l-0439). The authors are grateful for support from an MOE ACRF Tier-1 NUS startup grant. The authors gratefully acknowledge advice, expertise and fabrication support from the IME team, including in particular Patrick Lo Guo-Qiang, Andy Lim Eu-Jin. C.G. acknowledges support from the Swiss National Scientific Foundation through an Ambizione Fellowship. N.H. acknowledges that this material is based upon work supported by the National Science Foundation Graduate Research Fellowship under Grant No. 1122374. J.M. acknowledges support from the iQuISE fellowship. The authors would like to thank Mihir Pant for his commentary on the manuscript.


\end{document}